\documentclass[11pt]{article}
\newcommand{\Comment}[1]{{}}
\usepackage[textwidth = 430 pt, textheight = 630 pt]{geometry}
\usepackage{amsmath,euscript,amssymb,amsfonts,graphicx,bbm}

\usepackage{color}
\usepackage{bbold}
\definecolor{MyDarkBlue}{rgb}{0.15,0.15,0.45}
\usepackage[linktocpage=true]{hyperref}
\hypersetup{
colorlinks=true,
citecolor=MyDarkBlue,
linkcolor=MyDarkBlue,
urlcolor=MyDarkBlue,
pdfauthor={Constantinos Papageorgakis and Andrew~B.~ Royston},
pdftitle={Scalar Soliton Quantization with Generic Moduli},
pdfsubject={hep-th}
}

\usepackage[numbers,sort&compress]{natbib}
\usepackage{hypernat}
\usepackage{slashed}

\newcommand\ignore[1]{}
\def\half{\frac{1}{2}}

\def\NN{\mathcal{N}}

\def\bx{\mathbf{x}}
\def\by{\mathbf{y}}
\def\bz{\mathbf{z}}
\def\bw{\mathbf{w}}

\def\bk{\mathbf{k}}

\def\bkp{\mathbf{k'}}

\def\ed{\, \textrm{d}}

\def\one{{\,\hbox{1\kern-.8mm l}}}

\def\hPhi{\hat{\Phi}}
\def\hPi{\hat{\Pi}}
\def\hU{\hat{U}}
\def\hp{\hat{p}}
\def\hchi{\hat{\chi}}
\def\hpi{\hat{\pi}}
\def\ha{\hat{a}}
\def\habar{\hat{\bar{a}}}

 \newcommand{\pd}{\partial}

\newcommand{\Cset}{{\,\,{{{^{_{\pmb{\mid}}}}\kern-.45em{\mathrm C}}}}}
     \newcommand{\cI}{\mathcal I} \newcommand{\cJ}{\mathcal
  J}        
 
\newcommand{\ie}{{\it i.e.~}} \newcommand{\eg}{{\it e.g.~}}
 \newcommand{\etal}{{\it et al.~}}
\newcommand{\be}{\begin{equation}} \newcommand{\ee}{\end{equation}}
\newcommand{\bea}{\begin{eqnarray}} \newcommand{\eea}{\end{eqnarray}}

 \newcommand{\OO}{\mathcal{O}}
\newcommand{\MM}{\mathcal{M}}

\begin{document}

 \rightline{RUNHETC-2014-02}
 \rightline{QMUL-PH-14-05}
 \rightline{MIFPA-14-08}

   \vspace{2truecm}

\centerline{\LARGE \bf {\sc Scalar Soliton Quantization}} 
\vspace{.5cm}\centerline{\LARGE \bf {\sc with Generic Moduli}} 
\vspace{1.5truecm}
\centerline{ {\large {\bf{\sc Constantinos
        Papageorgakis}${}^{\,a,b,}$}}\footnote{E-mail address:
    \href{mailto:Costis Papageorgakis
      <c.papageorgakis@qmul.ac.uk>}{\tt
      c.papageorgakis@qmul.ac.uk}} {and} {\large {\bf{\sc
        Andrew~B.~Royston}${}^{\,c,}$}}\footnote{E-mail address:
    \href{mailto:Andy Royston <aroyston@physics.tamu.edu>}{\tt
      aroyston@physics.tamu.edu} } }

\vspace{1cm}
\centerline{${}^a${\it  NHETC and Department of Physics and Astronomy}}
\centerline{{\it  Rutgers University, Piscataway, NJ 08854-8019, USA}}
\vspace{.5cm}
\centerline{${}^b${\it CRST and School of Physics and Astronomy}}
\centerline{{\it Queen Mary, University of London, E1 4NS, UK}}
\vspace{.5cm}
\centerline{${}^c${\it George~P.~\& Cynthia Woods Mitchell Institute}}
\centerline{{\it for Fundamental Physics and Astronomy}}
\centerline{{\it Texas A\&M University, College Station, TX 77843, USA}}

\vspace{1.5truecm}

\thispagestyle{empty}

\centerline{\sc Abstract}
\vspace{0.4truecm}
\begin{center}
  \begin{minipage}[c]{380pt}{We canonically quantize multi-component
      scalar field theories in the presence of solitons. This extends
      results of Tomboulis \cite{Tomboulis:1975gf} to general soliton
      moduli spaces. We derive the quantum Hamiltonian, discuss
      reparameterization invariance and explicitly show how, in the
      semiclassical approximation, the dynamics of the full theory
      reduce to quantum mechanics on the soliton moduli space.  We
      emphasize the difference between the semiclassical approximation
      and a truncation of the dynamical variables to moduli.  Both
      procedures produce quantum mechanics on moduli space, but the
      two Hamiltonians are generically different.}
\end{minipage}
\end{center}

\vspace{.4truecm}

\noindent

\vspace{.5cm}

\setcounter{page}{0}

\newpage

\renewcommand{\thefootnote}{\arabic{footnote}}
\setcounter{footnote}{0}

\linespread{1.1}
\parskip 7pt

{}~
{}~

\makeatletter
\@addtoreset{equation}{section}
\makeatother
\renewcommand{\theequation}{\thesection.\arabic{equation}}

\section{Introduction and Summary}

The topic of quantization in soliton sectors is a rich one
with a long list of applications. Foundational work on this subject
was carried out in the mid 70's and includes \cite{Dashen:1974ci,
  Dashen:1974cj, Goldstone:1974gf, Gervais:1974dc, Tomboulis:1975gf,
  Callan:1975yy, Christ:1975wt, Gervais:1975pa,Tomboulis:1975qt}. For
thoroughly pedagogical reviews we refer the reader to
\cite{Jackiw:1977yn, Rajaraman:1982is}. 

In \cite{Tomboulis:1975gf} Tomboulis quantized a simple
two-dimensional scalar theory in the one-soliton sector by introducing
a canonical transformation from the original fields to a dynamical
modulus---\ie a collective coordinate---plus fluctuations around the
classical soliton solution, while imposing a set of constraints which
preserves the total number of degrees of freedom. A key assumption in
that work was that the soliton solution has a single modulus
associated with translations in the spatial direction, as is the case
\eg for a kink in $\phi^4$ theory. This greatly simplifies some
conceptual and calculational aspects of the analysis.  Similarly,
other contemporary approaches to soliton quantization around static
classical solutions, including canonical transformations with
unconstrained variables as well as path integral techniques, primarily
dealt with systems in which all collective coordinate degrees of
freedom correspond to translational modes.\footnote{Some aspects of
  the analysis of \cite{Christ:1975wt}, in particular the derivation
  of the soliton sector Hamiltonian, are more general and do not
  require the linear motion assumption for the collective coordinates
  made elsewhere in the paper---an assumption which is based on an
  identification of the collective coordinates with translational
  degrees of freedom.}

Several years later, a more geometrical framework for understanding
collective coordinates emerged from studies of the Bogomolny equation,
describing `t Hooft--Polyakov monopoles \cite{'tHooft:1974qc,
  Polyakov:1974ek} in four-dimensional Yang--Mills--Higgs theory
in the BPS limit of vanishing potential \cite{Bogomolny:1975de,
  Prasad:1975kr}.  In this theory, the minimal-energy solution set of
the static field equations with fixed boundary conditions,
corresponding to a particular topological charge sector, is a
finite-dimensional Riemannian manifold, $(\MM, G)$.  The metric is the
natural one induced from the (flat) metric on field configuration
space.  The framework suggested by Manton \cite{Manton:1981mp} is
that, for slowly varying field configurations, the dynamics of the
full system is well approximated by promoting the moduli to
time-dependent variables---the collective coordinates---in which case
the field theory equations of motion reduce to the geodesic equation
on $\MM$.  The usefulness of this framework was beautifully
demonstrated by Atiyah and Hitchin's analysis of two-to-two monopole
scattering \cite{Atiyah:1988jp}.

Generically, $\MM$ has curvature and not all moduli correspond to
broken symmetries such as translations.  Nevertheless in asymptotic
regions of $\MM$, corresponding to field configurations with
well-separated and localized lumps of energy, one can associate the
parameters with locations and internal phases of constituent solitons.
This suggests that Manton's paradigm of motion on moduli space is
applicable in any theory admitting static multi-soliton solutions.

It should be emphasized that Manton's prescription is for constructing
approximate time-dependent solutions to \emph{classical} field
equations.  However, as demonstrated earlier by Gervais, Jevicki, and
Sakita \cite{Gervais:1975pa}, it is also natural to assume small
velocities for the collective coordinates in the semiclassical
analysis of a soliton sector of a quantum theory. The approximate
classical solution provides an approximate saddle point for the
semiclassical expansion of the path integral. One would like the
corrections coming from performing the saddle-point approximation to
be comparable to those due to expanding around an approximate
solution; the latter are controlled by the collective coordinate
velocities.\footnote{In the rare circumstance where the time-dependent
  classical solution is exact, one can employ the more powerful method
  of \cite{Dashen:1974ci}, which takes the form of a WKB
  approximation; see \eg \cite{Dashen:1975hd}.  Our focus here will be
  on the semiclassical expansion around static soliton solutions,
  since this is typically all one has to work with in going beyond the
  two-dimensional kink.}  Hence the geometry $(\MM, G)$ provides a
natural starting point for the quantum analysis of a soliton sector,
and the Manton approximation is incorporated as part of the
semiclassical expansion.

This point of view was first considered in \cite{Gibbons:1986df} and
has since been used to great effect, \eg in the context of $\NN = 2$
supersymmetric four-dimensional gauge theory \cite{Sethi:1995zm,
  Cederwall:1995bc, Gauntlett:1995fu} where semiclassical results can
be compared against the quantum-exact ones of Seiberg and Witten
\cite{Seiberg:1994rs,Seiberg:1994aj}. In these analyses one typically
truncates the classical degrees of freedom to the collective
coordinates and then quantizes the resulting finite-dimensional
system, yielding a (supersymmetric) quantum mechanical sigma model
with target $\MM$.  This is sufficient for answering basic questions
about the original quantum field theory, such as the existence of
soliton states and what charges these carry.  The first corrections to
masses and charges, obtained from one-loop determinants, have also
been considered \cite{Kaul:1984bp,Rebhan:2006fg}.  However, to our
knowledge, the exact quantum Hamiltonian describing the full dynamics
of a theory around a (multi-) soliton sector has not been studied
within the general geometrical framework.\footnote{See
  \cite{Fujii:1997pt} for an interesting, if somewhat implicit,
  construction from the point of view of embedded submanifolds.}

In this work we extend the canonical transformation of
\cite{Tomboulis:1975gf} to multi-component scalar field theories with
general multi-soliton moduli spaces. This naturally requires using
geometric quantities on the moduli space of classical solutions. Our
primary goal is to extract the quantum Hamiltonian for this system,
which may be useful in various contexts, such as the study of
scattering processes involving both solitons and perturbative
particles \cite{Papageorgakis:2014dma}.

Our secondary goal is to establish a formalism that facilitates
extending this inquiry to (supersymmetric) theories with gauge fields
and fermions, \eg involving monopoles in four dimensions or
instanton-solitons in five dimensions.  While we intend to return to
this in the near future, the restriction to scalar fields helps
highlight the main qualitative results against the added technical
details required for those applications.

In that vein, we emphasize several conceptual points as they arise in
the explicit analysis.  We demonstrate how one recovers a
reparameterization-invariant theory for the dynamical moduli when the
fluctuations are switched off. This sector has knowledge of both the
intrinsic and extrinsic geometry of the system. Furthermore, we show
how the full quantum Hamiltonian of the field theory can be expanded
in the perturbative coupling, when one additionally requires the
solitons to be slowly moving. We organize and present this
semiclassical expansion to the first few orders and briefly discuss
how Lorentz invariance can be recovered in perturbation theory.
Finally, we exhibit how, when restricting to incoming and outgoing
states which do not involve perturbative excitations, the
leading-order dynamics reduce to quantum mechanics on the soliton
moduli space.

The rest of this paper is organized as follows: In Sect.~\ref{change}
we set up the background, introduce the change of variables and
canonically quantize the theory. In Sect.~\ref{solitonham} we obtain
the quantum Hamiltonian. Sect.~\ref{covariance} deals with the
reparameterization invariance of the collective coordinate sector,
while in Sect.~\ref{semianalysis} we present the semiclassical
expansion. Finally, Sect.~\ref{reduction} discusses the reduction to
quantum mechanics on the moduli space.

\section{The change of variables}\label{change}

We begin with a general class of real scalar field theories with
classical Lagrangian
\begin{equation}\label{lagrangian}
L = \int \ed\bx \left\{ \half \dot{\Phi} \cdot \dot{\Phi}
  -\half\pd_\bx \Phi \cdot \pd_\bx \Phi - V(\Phi) \right\}~.
\end{equation}
We work in flat $D$-dimensional Minkowski space, with $\bx$ a
$(D-1)$-dimensional position vector and $\ed\bx$ shorthand for
$\ed^{D-1} x$. $\Phi$ is an $n$-tuple and $\cdot$ denotes the
Euclidean inner product on $\mathbb{R}^n$.  When necessary we will use
indices $a,b,...$ to label components of $n$-tuples.  Let $M_{\rm vac}
= \{ \Phi ~|~ V(\Phi) = 0 \} \subset \mathbb{R}^n$ denote the space of
vacua where the potential energy function vanishes.  A finite-energy
field configuration must approach some point in $M_{\rm vac}$ as $\bx
\to \infty$ in any direction.  Thus the space of static, finite-energy
field configurations decomposes into topological sectors labeled by
$\pi_{D-2}(M_{\rm vac})$, the set of homotopy equivalence classes of
maps from the $(D-2)$-sphere at spatial infinity into the vacuum
manifold.  A (multi-) soliton solution\footnote{Although Derrick's
  theorem \cite{Derrick:1964ww,Manton:2004tk} precludes the existence
  of soliton solutions for $D > 2$, it is no more difficult to leave
  $D$ arbitrary. Doing so will facilitate the extension to theories
  with gauge interactions where one can have $D > 2$.} will be a field
configuration of minimal energy in a nontrivial topological sector.
In particular, $M_{\rm vac}$ should have multiple components in order
for solitons to exist when $D= 2$.

The Hamiltonian, $H[\Phi,\Pi]$ associated with the Lagrangian
$L[\Phi,\dot{\Phi}]$ is
\begin{equation}
  \label{eq:hamiltonian}
  H = \int \ed{\bf x}\; \Big[ \half \Pi\cdot \Pi + \half\pd_\bx \Phi
  \cdot \pd_\bx \Phi+  V(\Phi)\Big] ~.
\end{equation}
We assume that $\Phi,\Pi$ at fixed time $t$ are Darboux coordinates on
phase space
\begin{eqnarray}\label{flatps}
 \{ \Phi^a(t,\bx), \Phi^b(t,\by) \} &=& \{ \Pi^a(t,\bx), \Pi^b(t,\by)
\} \;\;= \;\; 0 \cr 
 \{ \Phi^a(t,{\bf x}), \Pi^b(t,{\bf y}) \} &=& \delta^{ab} \delta({\bf
  x} - {\bf y})~, 
\end{eqnarray}
where $\delta(\bx - \by)$ is a $(D-1)$-dimensional Dirac delta function and 
the Poisson bracket is given by
\begin{equation}
\{ F[\Phi,\Pi], \tilde F[\Phi,\Pi] \} := \int \ed\bz \left\{ \frac{\delta
    F}{\delta \Phi({\bf z})} \cdot \frac{\delta  \tilde F}{\delta \Pi{(\bf z)}}
  -\frac{\delta F}{\delta \Pi({\bf z})} \cdot \frac{\delta  \tilde F}{\delta
    \Phi{(\bf z)}}  \right\}~. 
\end{equation}
In the quantum theory, $\Phi,\Pi$ are promoted to
operators\footnote{For added clarity in this section we use hats to
  distinguish quantum operators from their classical counterparts.  In
  later sections we will be working exclusively at the quantum level
  and will drop this convention in favor of brevity.} $\hat{\Phi},
\hat{\Pi}$ and the Poisson bracket to a commutator
\begin{equation}
\{~,~\} \to [~,~] = i \{~,~\}~,
\end{equation}
such that
\begin{equation}
[ \hPhi^a(t,{\bf x}), \hPi^b(t,{\bf y}) ] = i \delta^{ab}
\delta^{(D-1)}(\bx - \by)~. 
\end{equation}

We consider a fixed topological sector and assume there exists a
finite-dimensional smooth family of classical static soliton
solutions, parameterized by moduli $U^M$,
\begin{equation}\label{soliton}
\Phi(x) = \phi({\bf x};U^M)~,
\end{equation}
where $M$ runs over the dimension of the moduli space
$\textrm{dim}_{\mathbb R}\mathcal M $, such that
\begin{equation}\label{Deltadef}
- \pd_{\bx}^2\phi +\frac{\delta V}{\delta \Phi} \bigg|_{\Phi = \phi} = 0~, \qquad
- \pd_{\bx}^2 + \frac{\delta^2 V}{\delta \Phi \delta \Phi}
\bigg|_{\Phi = \phi} =:
\Delta(U) \geq 0~. 
\end{equation}
The inequality $\Delta(U) \geq 0$ is meant to signify that $\Delta(U)$
is a positive operator, such that all of its eigenvalues are
non-negative, and the notation is to emphasize that this operator
depends on where we are in moduli space.

In order to study the behavior of the theory around the soliton
configuration \eqref{soliton}, one makes a change of variables from
the original field $\Phi(x)$ to collective coordinates $U^M =
U^M(t)$ and fluctuations $\chi(x;U^M(t))$ about the solution:
\begin{equation}\label{cov}
\Phi(x) = \phi(\bx;U^M(t)) + \chi(x;U^M(t))~.
\end{equation}
To preserve the number of degrees of freedom, there should be as many
constraints on these new variables as there are coordinates $U^M$.
We note that $\pd_M \phi$ will be a zero-mode of the linear
differential operator $\Delta$. One would like to exclude such
zero-frequency modes from the mode expansion of $\chi$.  This can be
done by imposing the constraints
\begin{equation}\label{chiconstraint}
\psi_{M}^{(1)} = \int \ed\bx\; \chi\cdot \pd_M \phi = 0~.
\end{equation}

We introduce momentum variables $(p_M, \pi(x;U^M))$ conjugate to
$(U^M,\chi)$ and extend this transformation to phase space.  We treat
these as Darboux coordinates in an extended phase space
\begin{equation}
  \{U^M(t),p_N(t)\}' = {\delta^M}_N~, \qquad \{ \chi^a(t,\bx;U(t)),
  \pi^b(t,\by;U(t)) 
  \}' = \delta^{ab} \delta(\bx - \by)~,  
\end{equation}
with Poisson structure $\{~,~\}'$, defined by 
\begin{align}
  \{ F[U,\chi;p,\pi],\tilde{F}[U,\chi;p,\pi] \}' :=&~ \frac{\pd F}{\pd
    U^M} \cdot \frac{\pd \tilde{F}}{\pd p_M} - \frac{\pd F}{\pd p_M}
  \cdot \frac{\pd \tilde{F}}{\pd U^M} + \cr & + \int \ed\bz \left\{
    \frac{\delta F}{\delta \chi({\bf z})} \cdot \frac{\delta
      \tilde{F}}{\delta \pi{(\bf z)}} -\frac{\delta F}{\delta \pi({\bf
        z})} \cdot \frac{\delta \tilde{F}}{\delta \chi{(\bf z)}}
  \right\}~.
\end{align}
We also extend the coordinate transformation \eqref{cov} to a phase
space transformation with the ansatz
\begin{equation}\label{cov2}
\Pi(x) = \Pi_{0}^M[U,\chi;p_M, \pi]\; \pd_M \phi(\bx;U(t)) + \pi(x,U(t))~,
\end{equation}
where the functionals $\Pi_{0}^M$ will be determined below.  In
analogy with \eqref{chiconstraint} we impose
\begin{equation}\label{piconstraint}
\psi_{M}^{(2)} = \int \ed\bx \;\pi\cdot \pd_M \phi = 0~.
\end{equation}

The constraints are second-class as the Poisson brackets are non-vanishing:
\begin{eqnarray}
  \{ \psi_{M}^{(1)}, \psi_{N}^{(1)} \}' &=& \{ \psi_{M}^{(2)},
  \psi_{N}^{(2)} \}' = 0~,\cr \{ \psi_{M}^{(1)}, \psi_{N}^{(2)} \}'
 & =& \int \ed\bz \;\pd_M \phi \cdot\pd_N \phi =: G_{MN}(U)~.
\end{eqnarray}
Here $G_{MN}(U)$ is the metric on the moduli space of soliton
solutions.  Restriction of the dynamics to the constraint surface is
achieved through the introduction of Dirac brackets,
\begin{equation}
\{ F, \tilde{F} \}_{\rm D}' := \{ F, \tilde{F} \}' + \{ F,
\psi_{M}^{(1)} \}' G^{MN} \{ \psi_{N}^{(2)}, \tilde{F} \}'  -  \{ F,
\psi_{M}^{(2)} \}' G^{MN} \{ \psi_{N}^{(1)}, \tilde{F} \}' ~. 
\end{equation}
Geometrically, the Dirac bracket is the pullback of the Poisson
bracket to the constraint surface and satisfies all the properties of
the ordinary Poisson bracket.  The appearance of the moduli space
metric $G_{MN}$ in the Dirac bracket is quite natural and can be
viewed as a motivation for choosing the momentum constraint as in
\eqref{piconstraint}. 

One can straightforwardly work out the Dirac brackets of our Darboux
coordinates.  The nonzero brackets with the constraints are
\begin{eqnarray}
  \{\psi_{N}^{(1,2)},p_M \}' &=& \pd_M \psi_{N}^{(1,2)}\cr
  \{ \chi,\psi_{M}^{(2)} \}' &=& \pd_M \phi\cr
  \{\psi_{M}^{(1)}, \pi \}' &=& \pd_M \phi\;,
\end{eqnarray}
so that we have
\begin{eqnarray}\label{DiracBrackets}
 \{ U^M, p_N \}_{\rm D}' &=& {\delta^M}_N~, \cr
 \{ p_M, p_N \}_{\rm D}' &=& - ( \pd_M \psi_{P}^{(1)} ) G^{PQ} (\pd_N
\psi_{Q}^{(2)}) + ( \pd_M \psi_{P}^{(2)} ) G^{PQ} (\pd_N
\psi_{Q}^{(1)})~, \cr 
 \{p_M, \chi(\bx) \}_{\rm D}' &=& - \pd_M \chi(\bx) + (\pd_M \psi_{P}^{(1)} )
G^{PQ} \pd_Q \phi(\bx)~, \cr 
 \{p_M, \pi(\bx) \}_{\rm D}' &=& - \pd_M \pi(\bx) + (\pd_M \psi_{P}^{(2)} )
G^{PQ} \pd_Q \phi(\bx)~, \cr 
 \{ \chi^a(\bx), \pi^b(\by) \}_{\rm D}' &=& \delta^{ab} \delta(\bx -
 \by) - \pd_M 
\phi^a(\bx) G^{MN} \pd_N \phi^b(\by)~, 
\end{eqnarray}
with the rest vanishing.  Here we suppressed all non-essential
arguments of the fields.  These brackets appear complicated at first,
but we have not yet specified the functional dependence of $\chi,\pi$
on $U^M$. One can freely do this, since the degrees of freedom
contained in $\chi$ should comprise a basis for
$L^2[\mathbb{R}^{D-1}]$ and not $L^2[\mathbb{R}^{D-1} \times
\MM]$. Indeed, it is always possible to choose the $U$-dependence of
$\chi,\pi$ such that
\begin{equation}\label{Uconstraint}
\pd_M \psi_{N}^{(1,2)} \approx 0~,
\end{equation}
where $\approx$ denotes `upon restriction to the constraint surface';
see App.~\ref{app:modes} for details. Having done so, the
non-vanishing Dirac brackets become
\begin{eqnarray}\label{DBs2} 
  \{ U^M, p_N \}_{\rm D}' &=& {\delta^M}_N\cr  \{p_M, \chi(\bx)
  \}_{\rm D}' &\approx& - \pd_M \chi(\bx)\cr  \{p_M, \pi(\bx) \}_{\rm
    D}' &\approx& - 
  \pd_M \pi(\bx) \cr  \{ \chi^a(\bx), \pi^b(\by) \}_{\rm D}' &=&
  \delta^{ab} \delta(\bx - \by) - \pd_M \phi^a(\bx) G^{MN} \pd_N
  \phi^b(\by)~. \raisetag{12pt}
\end{eqnarray}
We remind that for systems with second-class constraints it is the
Dirac bracket that is promoted to the commutator in the quantum theory
\begin{equation}
\{~,~\}_{\rm D}' \to [~,~]' = i \{~,~\}_{\rm D}'~.
\end{equation}

In order for the quantum theory in the old and new variables to be
equivalent, we must require that the transformation $(\Phi;\Pi) \to
(U^M,\chi; p_M,\pi)$ defined by \eqref{cov} and \eqref{cov2} be
canonical. Then $\{~,~\} = \{~,~\}_{\rm D}'$ and hence $[~,~] =
[~,~]'$.\footnote{In particular, the restriction of the new extended
  phase space to the constraint surface should give back the original
  phase space.}  This latter condition can be used to fix the
functionals $\Pi_{0}^M$ in \eqref{cov2}.  In order to implement this
requirement we compute $\{ \Phi^a(x), \Phi^b(y) \}_{\rm D}'$, $\{
\Phi^a(x), \Pi^b(y) \}_{\rm D}'$, and $\{ \Pi^a(x), \Pi^b(y) \}_{\rm
  D}'$ by inserting the change of variables \eqref{cov}, \eqref{cov2}
and using the brackets \eqref{DBs2}.  In the process, we find
$\Pi_{0}^M$ such that the results are consistent with
\eqref{flatps}. The full computations are tedious but straightforward;
some intermediate results are recorded in App.~\ref{app:commutators}
for the reader interested in the derivation.

We summarize these results as follows. The $[\hPhi,\hPhi]$ commutator
is trivial since $\hU,\hchi$ are commuting operators.  Thus
\begin{equation} 
[\hPhi^a(\bx), \hPhi^b(\by)]' = [ \phi^a(\bx;\hU) + \hchi^a(\bx;\hU),
\phi^b(\by;\hU) + \hchi^b(\by;\hU) ]' = 0~. 
\end{equation}
The calculation of $[\hPhi,\hPi]'$ fixes the form of $\Pi_{0}^M$.  At
the classical level one finds
\begin{equation}\label{classicalPi0}
\Pi_{0}^N \approx \left (p_M - \int \ed\bz \pi(\bz;U) \cdot\pd_M
  \chi(\bz;U) \right) [ (G - \Xi)^{-1} ]^{MN}~, 
\end{equation}
where
\begin{equation}\label{Xidef}
\Xi_{MN}(U) := \int \ed\bz  \chi(\bz;U) \cdot \pd_M \pd_N
\phi(\bz;U)~. 
\end{equation}

At the quantum level one must be careful about
operator ordering in Eq.~\eqref{cov2}. The symmetrized ansatz
\begin{equation}\label{quantumPi0}
\hPi(x) = \half \left( \ha^M \pd_M \phi(\bx; \hU(t)) + (\pd_M
  \phi(\bx;\hU(t))) \habar^M \right) + \hpi(x;\hU(t))~, 
\end{equation}
where
\begin{align}\label{as}
\ha^M :=&~ \left( \hp_N - \smallint \hpi \cdot \pd_N \hchi \right) [
(\hat{G} - \hat{\Xi})^{-1} ]^{NM} \cr
\habar^M :=&~  [ (\hat{G} - \hat{\Xi})^{-1}]^{MN} \left(\hp_N -
  \smallint \pd_N \hchi \cdot \hpi \right)~, 
\end{align}
provides a natural generalization of the ansatz in
\cite{Tomboulis:1975gf}.  Here we have begun using the shorthand $\int
\ed\bz \pi(\bz;U)\cdot \pd_N \chi(\bz;U) = \int \pi \cdot \pd_N \chi$.
It is also useful to introduce the combination
\begin{equation}
\hat C^{MN} := [(\hat G - \hat \Xi)^{-1}]^{MN}~.
\end{equation}
Note that $\hat C^{MN} = \hat C^{(MN)}$ and that \eqref{quantumPi0}
reduces to \eqref{classicalPi0} when the operators become commuting
fields.  With the form of the change of momentum variables fixed, it
is now a nontrivial task to check whether $[\hPi,\hPi]' = 0$.
Explicit evaluation leads to the expected result.\footnote{In place of
  \eqref{cov} one could have also used an alternative change of
  variables as in \cite{Christ:1975wt}---see also
  \cite{Fujii:1997pt}--- where the fluctuation field $\chi$ can be
  directly expanded in terms of only non-zero-modes. Then one does not
  need to impose constraints and the $p$ plus $\pi$-modes are canonically
  conjugate to the $U$ plus $\chi$-modes. It can be explicitly seen that this
  approach also leads to the relation \eqref{classicalPi0} and hence
  the same soliton sector Hamiltonian.}

\section{The soliton sector Hamiltonian}\label{solitonham}

We are now in a position to implement the change of variables
\eqref{cov} and \eqref{quantumPi0} in the Hamiltonian
\eqref{eq:hamiltonian}. Squaring \eqref{quantumPi0} leads
to\footnote{From now on we will drop hats as well as the $\approx$ notation,
  since all expressions are understood as restricted to the constraint
  surface.}
\begin{align}\label{Pisq}
  \int \ed\bx\; \Pi \cdot \Pi =&~ A^M G_{MN} A^N + \int \pi \cdot \pi-
  \frac{1}{4} C^{MP} C^{NQ} \int \pd_M \pd_P \phi \cdot \pd_N \pd_Q \phi\cr
  & + \frac{1}{2} C^{MP} C^{NQ} \Gamma_{MNR} C^{RS} \Big( \Gamma_{PQS}
  +2 \Gamma_{QSP}- \int \chi \cdot\pd_P \pd_Q \pd_S \phi\Big)\cr & -
  \frac{1}{2} C^{MP} C^{NQ} \pd_P \Gamma_{QMN} ~,
\end{align}
where
\begin{equation}
A^M := \half (a^M + \bar{a}^M)~.
\end{equation}
All terms beyond the first two result from the evaluation of two
commutators and should be thought of as $\OO(\hbar^2)$. We have also
introduced
\begin{equation}
\Gamma_{PMN} := \half \left( \pd_M G_{PN} + \pd_N G_{PM} - \pd_P
  G_{MN} \right) = \int \pd_P \phi \cdot\pd_M \pd_N \phi
\end{equation}
to define Christoffel symbols on the moduli space.

At this point we note that, in the special case where the moduli space
consists of a single modulus associated with translations, $\MM =
\mathbb{R}$, all terms in the second and third lines vanish. The terms
in the first line then reproduce the analogous result in
\cite{Tomboulis:1975gf}, including the `quantum correction' term
$\smallint (\pd^2 \phi)^2$.

The full Hamiltonian follows trivially from \eqref{Pisq} by adding the
potential, which can be expanded around the solution $\Phi = \phi$:
\begin{align}\label{QuantH}
H =&~ v(U) +  \half A^M G_{MN} A^N  
- \frac{1}{8} C^{MP} C^{NQ} \Big( \int \pd_M \pd_P \phi\cdot
  \pd_N \pd_Q \phi \Big) + \cr 
& + \frac{1}{4} C^{MP} C^{NQ} \Big[  - \pd_P \Gamma_{QMN} +
\Gamma_{MNR} C^{RS} \Big(  \Gamma_{PQS} +2 \Gamma_{QSP}- \int
\chi \cdot\pd_P \pd_Q \pd_S \phi\Big)   \Big]  \cr
& +\int \Big[\half
  \pi \cdot \pi + s(\bx;U) \cdot \chi + 
\frac{1}{2} \chi \cdot \Delta(\bx;U) \chi + V_I(\chi) \Big]~.
\end{align}
In the above
\begin{equation}
  v(U) :=  \int \Big( \half \pd_\bx \phi\cdot \pd_\bx
  \phi+V(\phi)\Big) ~, \qquad s(\bx;U) :=  -
    \pd_\bx^2 \phi+ \frac{
   \pd  V}{\pd \Phi} \bigg|_{\Phi = \phi}~
\end{equation}
and $V_I(\chi)$ denotes cubic and higher-order interaction terms in
the fluctuations $\chi$ coming from the original potential.  If
$\phi(\bx,U)$ parameterizes a family of {\it exact} static solutions
then: $a)$ $v(U)$ will be a constant, by definition the classical
soliton mass and $b)$ the source term $s(\bx;U)$ will vanish.
However, we will see shortly that it is natural to also allow for a
small deviation from an exact solution.  In that case $\MM$ is not
really a true moduli space, as evidenced by the appearance of the
potential $v(U)$.

Eq.~\eqref{QuantH} is the final, exact result for the quantum
Hamiltonian\footnote{\label{ft:ct}We have suppressed the appearance of
  counter-terms in our discussion, as they are model dependent.  A
  renormalizable theory will require a finite number of local
  counter-terms to be added to the action.  These counter-terms can be
  determined from the UV divergences in the perturbative sector of the
  theory.  The resulting counter-term Hamiltonian should also be
  transformed to the soliton sector and included in \eqref{QuantH}.
  It is a nontrivial test of renormalizability that the resulting
  counter-term Hamiltonian is sufficient to cancel all UV divergences
  for processes computed in the soliton sector.} of the theory. It is
valid for all values of soliton moduli $U^M$ and conjugate momenta
$p_N$.

\section{Covariance}\label{covariance}

Given the form of \eqref{QuantH}, it appears that the Hamiltonian is
not invariant under arbitrary reparameterizations of the moduli
$U$. This is not the case and the manifestly invariant form of the
Hamiltonian can be recovered once we properly order the kinetic term
operators.

The canonical change of variables in configuration space \eqref{cov}
from $\Phi$ to $U, \chi$---plus constraints---effectively maps a
Cartesian coordinate system to a curvilinear one. This map describes
how the curved moduli space is embedded in the total
infinite-dimensional space of modes.  The orthogonal directions to
$\MM$, within the constraint surface, are parameterized by the massive
oscillator modes of $\chi$.  It is possible to use the theory of
embedded surfaces in order to construct the exact metric for the
infinite-dimensional Cartesian space in the new curvilinear coordinate
system. This was explicitly done by Fujii \etal in
\cite{Fujii:1997pt}, who found that the Hamiltonian can be expressed
in terms of the Laplace-Beltrami operator in the curvilinear
coordinate frame. Hence the full theory, in the new set of variables,
is reparameterization invariant as expected.
 
It is interesting to see how covariance becomes manifest in the
subsector of the theory with all fluctuations switched off. We have,
from \eqref{QuantH},
\begin{align} \label{laplace} H|_{\chi,\pi =0} =&~ \half p_M
  G^{MN} p_N + v(U) +\frac{1}{8} (\pd_P G^{PM})G_{MN}(\pd_Q
  G^{QN}) -\frac{1}{4} \pd_M \pd_N G^{MN} +  \cr 
  &~ - \frac{1}{4} G^{MP}
  G^{NQ}\Big[\half\int \pd_M \pd_P \phi \pd_N \pd_Q \phi +\pd_P
  \Gamma_{QMN}\Big] + \cr
  &~ + \frac{1}{4}\Gamma^{PQS} \Big[ \Gamma_{PQS} + 2
  \Gamma_{QSP} \Big] ~,
\end{align}
where the last two terms in the first line are obtained from
commutators upon appropriately ordering the momentum operators in the
kinetic term. This expression can be manipulated as follows.  First note that
\begin{equation}\label{covtomb}
-\frac{1}{8} G^{MP} G^{NQ}\int \pd_M \pd_P \phi \pd_N \pd_Q \phi
=-\frac{1}{8} 
\int (\nabla^2 \phi)^2 -\frac{1}{8} {\Gamma_{RM}}^M {{\Gamma^R}_N}^N ~.
\end{equation}
Second, we have
\begin{equation}\label{divG}
\pd_M\pd_N G^{MN} =   4 Y -R + \Gamma_{RMS}\Gamma^{SMR} \;,
\end{equation}
where $R$ is the scalar curvature on moduli space
\begin{align}\label{Riemann1}
R =&~ G^{MP}G^{NQ}R_{MNPQ}  \cr
=&~ G^{MP}G^{NQ}\Big[\pd_N \Gamma_{QMP} - \pd_M
\Gamma_{QNP} + {\Gamma^{R}}_{NP} 
\Gamma_{RQM} - {\Gamma^{R}}_{MP} \Gamma_{RQN}\Big]
\end{align}
and \cite{Fujii:1997pt}
\begin{equation}
 Y := -\half \pd_M(G^{MN} {\Gamma^S}_{NS}) -\frac{1}{4}
  {\Gamma^{SN}}_S {\Gamma^R}_{NR} \;.
\end{equation}
Using this definition we can re-express \eqref{divG} as
\begin{align}\label{divGY}
-\frac{1}{4}  \pd_M\pd_N G^{MN} =&~   -\half Y +\frac{1}{4}R -\frac{1}{4}
\Gamma_{RMS}\Gamma^{SMR}  + \cr
&~ +  \frac{1}{4}\pd_M(G^{MN}G^{SR}\Gamma_{RNS})
+ \frac{1}{8}{\Gamma^{SN}}_M {\Gamma^R}_{NR}\;.
\end{align}
Substituting \eqref{covtomb} and \eqref{divGY} into \eqref{laplace},
one finds that all bilinears in the $\Gamma$'s as well as the terms
involving derivatives of $\Gamma$'s mutually cancel to leave
\begin{eqnarray}
  \label{laplacebeltrami}
H|_{\chi,\pi = 0}
&=&   \half p_M G^{MN} p_N + v(U)  -\frac{1}{2} Y +
\frac{1}{4}R  -\frac{1}{8} \int (\nabla^2 \phi)^2 \cr
&=&\half G^{-1/4} p_M G^{1/2} G^{MN} p_N G^{-1/4} + v(U)  +
\frac{1}{4}R  -\frac{1}{8} \int (\nabla^2 \phi)^2 \;.
\end{eqnarray}
In the last step we noted that
\begin{equation}\label{covariantK}
G^{-1/4} p_M G^{1/2} G^{MN} p_N G^{-1/4}  = p_M G^{MN}
  p_N  - Y \;. 
\end{equation}

The LHS of \eqref{covariantK} is in fact a covariant quantity when
understood as a Hamiltonian acting on a wavefunction $\Psi$ with
canonical normalization $\int_{\MM} \ed^d U \Psi^\ast \Psi = 1$.  The
time-independent Schr\"odinger equation takes the form
\begin{equation}\label{schr}
  \half G^{-1/4} \pd_M ( G^{1/2} G^{MN} \pd_N (G^{-1/4} \Psi )) = E \Psi\;,
\end{equation}
Redefining $\Psi = G^{1/4} \tilde \Psi$ leads to the
correct curved space normalization $\int_{\MM} \ed^d U \sqrt G \tilde\Psi^*
\tilde\Psi = 1$ and modifies \eqref{schr} to 
\begin{equation}
  \half G^{-1/2} \pd_M ( G^{1/2} G^{MN} \pd_N (\tilde\Psi )) = E \tilde\Psi\;,
\end{equation}
where the LHS is now the Laplace-Beltrami operator on the soliton moduli
space. 

Hence we have arrived at the explicitly covariant expression
\begin{equation}
  H|_{\chi,\pi = 0} = \half G^{-1/4} p_M G^{1/2} G^{MN} p_N G^{-1/4} + v(U) 
  +  \frac{1}{4}R  -\frac{1}{8} \int (\nabla^2 \phi)^2\;.
\end{equation}
In fact, the quantity $\smallint (\nabla^2\phi)^2$ has a nice
geometric interpretation.  Using the results of \cite{Fujii:1997pt}
one finds that\footnote{The notation of \cite{Fujii:1997pt} is rather
  different from the one used here, so it is useful to describe the
  precise map: We have $\pd_P\phi^a(\bx) \to B^{A x}_a$. Then
  $\nabla^2 \phi^a(\bx)\to g^{ab} \nabla_a B^{Ax}_b = n
  \overline{H}^{Ax}$ and the integration over $\bx$ is performed by
  contracting the $\overline H$'s with $\eta_{Ax,By}$.}
\begin{equation}
  \int (\nabla^2\phi)^2 = d^2 \mathcal H^2\;,
\end{equation}
where $\mathcal H$ is the extrinsic mean curvature and $d
=\textrm{dim}_{\mathbb R}\mathcal M $. This curvature invariant
encodes information about how the moduli space $\MM$ is embedded as a
submanifold into the infinite-dimensional flat configuration space.

Our covariant result is then simply
\begin{equation}\label{QMtruncate}
  H|_{\chi,\pi = 0} = \half G^{-1/4} p_M G^{1/2} G^{MN} p_N G^{-1/4} + v(U)
  + \frac{1}{4}R  -\frac{d^2}{8} \mathcal H^2 \;.
\end{equation}
This Hamiltonian defines a quantum mechanics with target $\MM$; we
will refer to \eqref{QMtruncate} as the `truncated Hamiltonian' in the
following.  The curvature terms are $\OO(\hbar^2)$ effects and may be
viewed as intrinsic and extrinsic `quantum potentials.'  The
appearance of the Ricci scalar is well documented in
background-independent approaches to quantum mechanics on curved
spaces.  In that context, it has been observed that the coefficient of
the Ricci scalar term is ambiguous, depending on the operator ordering
prescription \cite{DeWitt:1957at}.  Here there is no ambiguity because
the correct ordering prescription is inherited from the parent theory,
which is defined on a flat configuration space.  It is also
interesting to observe that even in the limit where the fluctuations
have been completely decoupled, the Hamiltonian for the collective
coordinates encodes information about the extrinsic geometry
\cite{Fujii:1997pt}.

\section{Semiclassical analysis}\label{semianalysis}

Up to this point we have not explicitly kept track of powers of
coupling constants.  As is typical in the soliton literature
\cite{Goldstone:1974gf,Jackiw:1977yn,Rajaraman:1982is}, we will assume
that there is effectively a single coupling $g$ such that, in terms of
the canonically normalized field $\tilde{\Phi}$, the potential
$\tilde{V}(\tilde{\Phi};g)$ has the scaling property
\begin{equation}
\tilde{V}(\tilde{\Phi};g) = \frac{1}{g^2} \tilde{V}(g \tilde{\Phi};1) =: \frac{1}{g^2}
V (g \tilde{\Phi})~. 
\end{equation}
Thus, if we define the rescaled field $\Phi = g \tilde{\Phi}$,
then then the entire coupling dependence of the
Lagrangian \eqref{lagrangian} becomes 
\begin{equation}\label{Lascale}
L(\tilde{\Phi},\dot{\tilde{\Phi}};g) = \frac{1}{g^2} L(\Phi, \dot{\Phi} ;1)~.
\end{equation}
We will assume that we have been working with the rescaled field
$\Phi$ all along and that we previously set the coefficient of
$g^{-2}$ in front of \eqref{lagrangian} to one.  Note that if $\phi$
is the rescaled classical solution, it will be independent of $g$ and
hence the canonically normalized classical solution $\tilde{\phi}$
will go as $g^{-1}$, which is the usual behavior we expect from a
soliton configuration.

Under the assumption \eqref{Lascale}, it is clear from the path
integral point of view that $g^2$ plays the role of $\hbar$ and the
semiclassical expansion is a $g$ expansion. Once the factor of
$g^{-2}$ is restored in front of the Lagrangian, the Hamiltonian,
\eqref{eq:hamiltonian}, becomes
\begin{equation}
  H = \int \ed \bx \left[ \frac{g^2}{2} \Pi \cdot \Pi + \frac{1}{g^2}
    \left( \half \pd_{\bx} \Phi \cdot \pd_{\bx} \Phi + V(\Phi) \right)
  \right]~. 
\end{equation}
Meanwhile, the definitions of the metric and potential on moduli space
read
\begin{equation}\label{resmetric}
  G_{MN} := \frac{1}{g^2}\int \pd_M \phi\cdot \pd_N \phi \;,\qquad
  \Xi_{MN} := \frac{1}{g^2}\int \chi\cdot \pd_M \pd_N \phi
\end{equation}
and
\begin{equation}
  v(U) :=   \frac{1}{g^2} \int \Big( \half \pd_\bx \phi\cdot \pd_\bx
  \phi+V(\phi)\Big) ~, \qquad s(\bx;U) :=   \frac{1}{g^2}\Big(-
  \pd_\bx^2 \phi+ \frac{
    \pd  V}{\pd \Phi} \bigg|_{\Phi = \phi}\Big)~\;.
\end{equation}
The canonical transformations are given by
\begin{eqnarray}\label{covrescaled}
\Phi &=& \phi + g \;\chi\cr
\Pi &=& \frac{1}{2}\Big( a^M \pd_M \phi +  \pd_M \phi \;\bar a ^M \Big) +
\frac{1}{g} \pi\;,
\end{eqnarray}
where 
\begin{equation}\label{newas}
  a^M = \frac{1}{g^2}(p_N  -  \smallint \pi \cdot\pd_N \chi) C^{MN} \;,\qquad
  \bar a^M = \frac{1}{g^2} C^{MN} (p_N  -  \smallint  \pd_N \chi\cdot \pi) \;,
\end{equation}
with $ C^{MN} = [(G - g \Xi)^{-1}]^{MN}$. In the above we have rescaled
the the fluctuations $\chi,\pi$ so that they are canonically
normalized fields, while the power of $g^{-2}$ in \eqref{newas} originates from
the definitions \eqref{resmetric}.

Setting $A^M = \half(a^M + \bar a^M) $ as before, our full quantum
Hamiltonian \eqref{QuantH} can now be re-written as
\begin{align}
H =&~   \frac{g^4}{2}  A^M
 G_{MN}  A^N  + v(U) 
- \frac{1}{8 g^2}  C^{MP} C^{NQ}  \int \pd_M
\pd_P   \phi \cdot \pd_N \pd_Q  \phi  \cr 
&  + \frac{1}{4}  C^{MP}  C^{NQ} \Big[  - \pd_P
 \Gamma_{QNM} +
\Gamma_{MNR}  C^{RS} \Big(  \Gamma_{PQS} +2 
\Gamma_{QSP}- \frac{1}{g}\int
 \chi \cdot\pd_P \pd_Q \pd_S  \phi\Big)   \Big]  \cr
& +\int \Big[\half
   \pi \cdot \pi + g\; s\cdot  \chi + 
\frac{1}{2}  \chi \cdot \Delta  \chi + V_I(
\chi) \Big] ~.
\end{align}
We can then expand the $ A^M G_{MN} A^N$ term in powers of the coupling as
follows: 
\begin{align}\label{AAexpansion}
 g^4  A^M G_{MN} A^N =&~ p_M \Big( G^{MN} + 2 g 
    (G^{-1} \Xi G^{-1})^{MN} + 3 g^2 (G^{-1} \Xi
    G^{-1} \Xi G^{-1})^{MN} + \OO(g^5) \Big) p_N \cr
&~ + \bigg[ \frac{1}{4} (\pd_P G^{PM})G_{MN}(\pd_Q G^{QN}) -
\frac{1}{2}\pd_M \pd_N G^{MN}  + \OO(g^3) \bigg]\cr
 & - \frac{1}{2} \bigg[   p_M  \Big( G^{MN}
   + 2 g (G^{-1} \Xi G^{-1})^{MN} + \OO(g^4)\Big) \int
 [\pd_N\chi^a,  \pi^a]_+ \cr 
& \qquad +  \int
 [\pi^a,  \pd_M\chi^a]_+ \Big( G^{MN}
   + 2 g (G^{-1} \Xi G^{-1})^{MN} + \OO(g^3)\Big) p_N\bigg]\cr
&~ +\frac{1}{4} \int [\pi^a,\pd_M\chi^a ]_+  \Big(
G^{MN}+ \OO(g^3)\Big) 
 \int [\pi^b,\pd_N \chi^b ]_+ ~,
\raisetag{24pt} 
\end{align}
where $[A,B]_+ := A B + B A$.  Note that, as in \eqref{laplace}, the
terms in the second line come from commutators when expanding $A^M
G_{MN}A^N$ and moving $p_M$ to the far left and $p_N$ to the far right
of the expression.

Notice also that the first line contains a term linear in the
fluctuations $\chi$ through $\Xi$.  The presence of this tadpole is
due to the fact that $\phi(\bx;U(t))$ is not an exact solution to the
time-dependent equations of motion, irrespective of whether or not
$\phi(\bx;U)$ is an exact solution to the time-independent ones. This
is what motivates the \emph{small velocity assumption}: As it stands,
\eqref{AAexpansion} is valid for all values of soliton momenta but
makes little sense in perturbation theory, since the scalar propagator
would be higher order in the coupling compared to the
tadpole. However, if one considers appropriately slowly-moving
solitons, $p^2 \chi$ can be viewed as a legitimate interaction term.

In a similar vein, since we do not solve the time-dependent equations
of motion exactly, there is no need to insist on an exact solution to
the time-independent equations.  We merely require an approximate
solution so that the tadpole term, $s(\bx;U) \cdot \chi$, coming from
the potential may also be viewed as an interaction term.

Thus we will continue by making the assumptions
\begin{equation}\label{qmanton}
  \dot U^M \sim \OO(g)\quad \Rightarrow \quad p_M \sim \OO(1/g)~, \qquad
  s(\bx;U) \sim \OO(1)~,  
\end{equation}
so that we are expanding around an approximate solution to the
time-dependent equations of motion.  Note that the latter condition
implies that
\begin{equation}\label{qmanton2}
v(U) =  M_{\rm cl} + 
  \delta v(U)~, \qquad \textrm{where} \quad  M_{\rm cl} \sim
  \OO(1/g^2)~, \quad  \delta v(U) \sim \OO(1)~.
\end{equation}
In other words, the integral of the potential evaluated on the
classical solution is constant up to $\OO(g^2)$-suppressed
corrections, which may be moduli dependent. The constant $M_{\rm cl}$
is interpreted as the classical---or leading order---contribution to the
soliton mass, while the corrections give a $U$-dependent potential on
the moduli space.

In this small-velocity and small-potential approximation, the
semiclassical expansion of the full Hamiltonian becomes
\begin{equation}\label{Hamexp}
H = H^{(-2)} + H^{(0)} + H^{(1)} + H^{(2)} + \OO(g^3)~,
\end{equation}
where
\begin{eqnarray}
  H^{(-2)} &=&  M_{\rm cl}~, \cr
  H^{(0)} &=& \half p_M G^{MN} p_N +  \delta
  v(U) + \half \int \left( \pi \cdot \pi + \chi \cdot \Delta \chi
  \right)~, \cr 
  H^{(1)} &=&  \int\bigg\{ \frac{1}{g}\; p_M G^{MP} \left( \chi \cdot
    \pd_P \pd_Q \phi \right) G^{PN} p_N + g\;s \cdot \chi +
  \frac{g}{3!} V^{(3)}_{abc}(\phi) \chi^a \chi^b \chi^c  \cr 
  && \qquad \quad - \frac{1}{4} \Big( [\pi^a, \pd_M \chi^a]_+ G^{MN}
    p_N + p_M G^{MN}[\pi^a, \pd_N \chi^a]_+
  \Big)  
  \bigg\}~, \cr 
  H^{(2)} &=& \frac{3g^2}{2} p_M \left( G^{-1} \Xi G^{-1} \Xi G^{-1}
  \right)^{MN} p_N + \frac{g^2}{4!}
  \int V^{(4)}_{abcd}(\phi) \chi^a \chi^b \chi^c \chi^d\cr
 && - \frac{g}{2}   \Big( [\pi^a, \pd_M \chi^a]_+   \left(
   G^{-1} \Xi G^{-1} \right)^{MN}  
    p_N + p_M   \left( G^{-1} \Xi G^{-1} \right)^{MN}  [\pi^a, \pd_N \chi^a]_+
  \Big)  \cr 
  && + \frac{1}{8} \left( \int [\pi^a,\pd_M\chi^a ]_+ \right)
  G^{MN} \left( \int [\pi^b,\pd_N \chi^b ]_+ \right)\cr
&&  +    \frac{1}{4} R - \half Y - \frac{1}{8g^2 } \int (\nabla 
    \phi)^2~.
\end{eqnarray}
Here, $H^{(n)}$ is $\OO(g^n)$ provided that \eqref{qmanton} and
\eqref{qmanton2} hold, and we recall that $G^{MN}\sim
\OO(g^2)$. $V^{(3,4)}(\phi)$ denote the third and fourth derivatives
of the potential, evaluated on the soliton solution $\phi$.  Finally,
we have used the results of Sect.~\ref{covariance} to simplify the
terms in $H^{(2)}$ that are zeroth order in fluctuations.

Let us briefly discuss the issue of Lorentz
invariance. Eq.~\eqref{Hamexp} is in principle a double expansion: a
quantum expansion in the coupling, as well as an expansion in small
soliton velocities.  A subset of the collective coordinates, $\{ U^i
\}_{i=1}^{D-1} \subset \{ U^M \}$, correspond to the center-of-mass
position of the soliton solution $\phi$. The conjugate variables,
$p_i$, correspond to the center-of-mass momentum.  One expects that
any observable computed exactly in the quantum theory should be
covariant under Lorentz transformations.  On the one hand, expanding
around slowly-moving solitons---in particular $p_i \sim
\OO(1/g)$---naturally breaks the Lorentz symmetry of the original
theory. On the other, the scaling \eqref{qmanton} suggests that
relativistic corrections should appear as quantum effects associated
with the $p_{i}^2\chi$ tadpoles. In fact, it can be explicitly seen
for the case of kink solitons in two-dimensional $\phi^4$ theory that
re-summing all the tree-level diagrams obtained by gluing together the
$p_{i}^2\chi$ tadpole interactions restores Lorentz invariance for the
soliton energy \cite{Gervais:1975pa, Gervais:1975yg}.  This
computation should be extendable to the class of theories we are
studying, but we will not explicitly consider it here.

\section{Reduction to QM on the soliton moduli space}\label{reduction}

It is straightforward to use our results for the semiclassical
expansion of the Hamiltonian to determine the behavior of the
leading-order dynamics. Keeping terms in $H$ through $\OO(1)$, we have
\begin{equation}\label{QMham}
 H =  M_{\rm cl} +  \half p_M G^{MN} p_N
+ \delta v(U)  + \half \int  \left( \pi \cdot \pi + \chi \cdot
  \Delta \chi \right) + \OO(g)\;.
\end{equation}
Let us focus on the fluctuation terms. We make a mode expansion
\begin{align}\label{modeexps}
  \chi(x;U) =&~ \int\frac{\ed\bf k}{(2\pi)^{D-1}}
  \frac{1}{\sqrt{2\omega_{\bf k}}} \Big[a_{\bf k}(t)  
  + a^\dagger_{-\bf k}(t) \Big] \zeta_{\bf k}(\bx;U)\cr
  \pi(x;U) =&~  \int\frac{\ed\bf k}{(2\pi)^{D-1}} (-i)
  \sqrt{\frac{\omega_{\bf k}}{ 2}} \Big[a_{\bf k}(t)  
 -  a^\dagger_{-\bf k}(t) \Big]\zeta_{\bf k}(\bx;U)\;,
\end{align}
where the $\zeta$'s are eigenfunctions of the operator $\Delta(U)$
with \emph{strictly positive} eigenvalues $\omega_{\bf k}^2$:
$\Delta(U) \zeta_{\bf k} = \omega^2_{\bf k}(U)\zeta_{\bf k}$. They are
orthonormal
\begin{equation}\label{ortho}
  \int\ed\bx \;\zeta_{\bf k}(\bx;U)\zeta_{\bf k'}(\bx;U) =
 (2\pi)^{D-1} \delta(\bf k-k')
\end{equation}
and satisfy the completeness relation
\begin{equation}\label{completeness}
  \int\frac{\ed\bf k}{(2\pi)^{D-1}} \zeta_{\bf k}(\bx;U)
  \zeta_{\bf k}(\by;U) =\delta(\bx-\by) - \frac{1}{g^2}
  \pd_M \phi(\bx;U)\cdot G^{MN} \pd_N \phi(\by;U)\; .
\end{equation}
The modified completeness relation is due to the fact that we have
excluded the zero-eigenvalue modes from the expansion.  The
$\zeta_{\bf k}(\bx,U)$ form a basis for the subspace of configuration
space orthogonal to the tangent space $T_U
\MM$.  Using
\eqref{completeness}, one can show that the commutator $[\chi,\pi]$,
\eqref{DBs2}, is equivalent to the standard creation and annihilation
commutators
\begin{equation}\label{creationops}
  [a_\bk, a_{\bkp} ] = [a^\dagger_\bk, a^\dagger_{\bkp} ]  = 0~,
  \qquad [a_\bk, a^\dagger_{\bkp} ] = (2\pi)^{D-1} \delta(\bk - \bkp)
  \;.  
\end{equation}

We have written the mode expansions \eqref{modeexps} as though the non-zero spectrum of $\Delta$ is purely continuous.  While the spectrum of $\Delta$ is guaranteed to have a continuous component,\footnote{\label{continuous} This statement can be justified as follows: Since classical solitons are localized objects, we expect
  the difference between the operator $\Delta(U)$ and the the operator
  $\Delta^0 := - \delta_{ab} \pd_{\bx}^2 +
  V_{ab}^{(2),\infty}(\hat{\bf x})$, to be a compact operator.  Here
  $V_{ab}^{(2),\infty}(\hat{\bf x}) \geq 0$ is the asymptotic form of
  the second derivative of the potential evaluated on the soliton
  solution as $\bx \to \infty$, and $\hat{\bx}$ parameterizes the
  $(D-2)$-sphere at infinity.  Weyl's theorem then implies that the
  continuous part of the spectra of $\Delta(U)$ and $\Delta^0$ must
  agree.  If $\min_{\hat{\bx}} V_{ab}^{(2),\infty}(\hat{\bf x}) > 0$,
  then the continuous spectrum of $\Delta^0$ will have a mass gap,
  while if $\min_{\hat{\bx}} V_{ab}^{(2),\infty}(\hat{\bf x}) = 0$ it
  will extend down to zero.  In either case there will be a continuous
  spectrum that we can label by ${\bf k}$.} there could additionally be a discrete component beyond the zero-modes.  Strictly positive discrete eigenvalues correspond to breather-like modes, and the mode expansion should include a sum over them.  We will understand  `$\int \ed\bk$' in the above and following
  expressions as representing the integral over the continuous
  spectrum plus the sum over the breather-like modes, if present.

Using \eqref{modeexps}, \eqref{ortho}, and \eqref{creationops}, it is
then easy to see that
\begin{equation}
   \half \int \ed\bx \left( \pi \cdot \pi + \chi \cdot  \Delta
     \chi\right) =  \int\frac{\ed\bk}{(2\pi)^{D-1}} \omega_\bk
   \Big(a_\bk^\dagger a_\bk + \frac{1}{2}[a_\bk,a^\dagger_\bk ]\Big)
\end{equation}
so that the full Hamiltonian is
\begin{equation}
 H \simeq  M_{\rm cl} + \half p_M G^{MN} p_N
+ \delta v(U) + \int\frac{\ed\bk}{(2\pi)^{D-1}} \omega_\bk
   \Big(\frac{1}{2}[a_\bk,a^\dagger_\bk] + a_\bk^\dagger a_\bk\Big)\;.  
\end{equation}

In particular, when acting on a state which does not involve massive
fluctuations, the last term above vanishes and one is left with the
zero-point energy of the fluctuation fields. In a renormalizable
theory, the divergent part of this quantity can be removed, after
vacuum energy subtraction, by mass renormalization; see
fn.~\ref{ft:ct}. The finite piece then generates a one-loop
correction to the potential $M_{\rm cl} + \delta v(U)\to
M_{\textrm{1-loop}} + \delta v(U)_{\textrm{1-loop}}$
\cite{Dashen:1974cj,Dashen:1974ci,Tomboulis:1975gf}. Hence, the final
result for the leading contribution in the semiclassical approximation
{\it and when restricting to incoming and outgoing states that do not
  contain perturbative excitations} is
\begin{equation}\label{QMsc}
H_{\rm s.c.} =  M_{\textrm{1-loop}}  + \half  p_M 
  G^{MN} p_N +  \delta v(U)_{\textrm{1-loop}}\;,
\end{equation}
which is a quantum mechanics on the soliton moduli space.  We will
refer to \eqref{QMsc} as the `semiclassical Hamiltonian'.

This quantum mechanics, as written, is not covariant with respect to
general coordinate transformations on $\MM$.  However, following the
discussion around \eqref{schr}, it can be trivially made covariant by
replacing $p_{M} G^{MN} p_N \to G^{-1/4} p_M G^{1/2} G^{MN} p_N
G^{-1/4}$.  These two quantities differ by $Y$, which is higher order
in the $g$-expansion and hence can be neglected in \eqref{QMsc}.

It is interesting to note that, even after this replacement, the two
quantum mechanical systems on $\MM$ defined by the truncated
Hamiltonian \eqref{QMtruncate} and the semiclassical Hamiltonian
\eqref{QMsc} are different.  Although the intrinsic and extrinsic
quantum potentials of the truncated Hamiltonian are present in the
semiclassical expansion \eqref{Hamexp}, it would be inconsistent to
include them in the semiclassical Hamiltonian \eqref{QMsc}, without
first accounting for all $\OO(g)$ and $\OO(g^2)$ corrections from
integrating out the fluctuations.  Furthermore, the semiclassical
approximation demands that the $\OO(1)$, `one-loop' corrections from
$\chi,\pi$ be accounted for in the semiclassical Hamiltonian: They are
of the same order as the kinetic term and moduli-dependent classical
potential $\delta v(U)$, due to the necessity of imposing
\eqref{qmanton} and \eqref{qmanton2}.

\newpage

\section*{Acknowledgments}

We would like to thank Sebastian Guttenberg for helpful discussions
and comments. CP is a Royal Society Research Fellow and partly
supported by the U.S. Department of Energy under grants DOE-SC0010008,
DOE-ARRA-SC0003883 and DOE-DE-SC0007897. ABR is supported by the
Mitchell Family Foundation. We would like to thank the Mitchell
Institute at Texas A\&M and the NHETC at Rutgers University
respectively for hospitality during the course of this work.  We would
also like to acknowledge the Aspen Center for Physics and NSF grant
1066293 for a stimulating research environment which led to questions
addressed in this paper.


\appendix

\section{Mode expansions for $\chi,\pi$}\label{app:modes}

For a fixed value of the moduli, $\chi$ and $\pi$ are simply
$n$-tuples of scalar fields on $\mathbb{R}^{D-1}$; they can be
expanded in any complete basis for the Hilbert space
$L^2[\mathbb{R}^{D-1},\mathbb{R}^n]$.  A particular basis that is
naturally adapted to the problem is the basis of eigenfunctions of the
Hermitian operator $\Delta(U)$, defined in \eqref{Deltadef}.  Since
the form of this operator depends on the moduli, so will its
eigenfunctions; we denote the complete set of eigenfunctions by $\{
\zeta_{\cI}(\bx;U) \}$, where ${\cI}$ runs over an indexing set.
This set will include both the continuous part of the
spectrum, as well as the discrete
part of the spectrum, which includes the zero-modes and may
additionally contain other massive breather-like modes.  We write
schematically
\begin{equation}\label{genmodeexp}
\chi(x;U) = \sum_{\cI} \chi^{\cI}(t) \zeta_{\cI}(\bx,U)~, \qquad
\pi(x;U) = \sum_{\cI} \pi^{\cI}(t) \zeta_{\cI}(\bx,U)~, 
\end{equation}
where $\chi^{\cI}(t), \pi^{\cI}(t)$ comprise the complete set of
degrees of freedom in $\chi(x;U),\pi(x;U)$.  The $\zeta_{\cI}$ satisfy
\begin{equation}\label{genorth}
  \int \ed \bx \zeta_{\cI}(\bx;U) \cdot \zeta_{\cJ}(\bx,U) =
  \delta_{\cI\cJ}~, \qquad \delta^{ab} \delta(\bx-\by) = \sum_{\cI}
  \zeta_{\cI}^a(\bx;U) \zeta_{\cI}^b(\by;U)~, 
\end{equation}
where by `$\delta_{\cI\cJ}$' and `$\sum_{\cI}$' we mean $(2\pi)^{D-1}
\delta(\bk - \bk')$ and $\int \frac{\ed \bk}{(2\pi)^{D-1}}$ in the
case of the continuous spectrum.

Let $e^A = {e^A}_M \ed U^M$ be a vielbein for the moduli space such
that $\delta_{AB} {e^A}_M {e^A}_N = G_{MN}$, where $\delta_{AB}$ is
the flat Euclidean metric on the tangent space, and let ${e_A}^M$
denote the inverse vielbein satisfying $\delta^{AB} {e_A}^M {e_B}^N =
G^{MN}$.  Then we know that the orthonormal eigenfunctions for the
zero-modes are
\begin{equation}
\zeta_{\cI=A}(\bx;U) = {e_A}^M(U) \pd_M \phi(\bx;U)~.
\end{equation}
Then, substituting \eqref{genmodeexp} into the constraints and using
\eqref{genorth}, we have
\begin{equation}
  \psi_{N}^{(1)} = \chi^A {e_A}^M(U) \int \pd_M \phi(\bx;U) \cdot \pd_N
  \phi(\bx;U) = \chi^A {e_A}^M(U) G_{MN}(U) = \chi_A {e^A}_N(U)~, 
\end{equation}
and similarly
\begin{equation}
\psi_{N}^{(2)} = \pi_A {e^A}_N(U)~.
\end{equation}
From here we can explicitly see that the constraint surface
corresponds to $\chi_A = \pi_A = 0$.  Meanwhile,
\begin{equation}\label{psivanish}
  \pd_M \psi_{N}^{(1)} = \chi_A \pd_M {e^A}_N(U) \approx 0~, \qquad
  \pd_M \psi_{N}^{(2)} = \pi_A \pd_M {e^A}_N(U) \approx 0~. 
\end{equation}

Another identity that follows trivially from \eqref{genmodeexp} and
\eqref{genorth} and will be useful below is
\begin{equation}\label{dchipi}
  \pd_M \left( \int \ed \bx \chi(\bx;U) \cdot \pi(\bx;U) \right) =
  \pd_M \left( \sum_I \chi_I \pi^I \right) = 0~, 
\end{equation}
or $\smallint \pd_M \chi \cdot \pi = - \smallint \chi \cdot \pd_M
\pi$.  Finally, if $I,J$ index the non-zero modes, which include the
continuous spectrum and any possible breather-like modes, then the
commutator of the fields $\chi,\pi$ is equivalent to
\begin{equation}
[\chi^I, \chi^J] = [\pi^I, \pi^J] = 0~, \qquad [\chi^I, \pi^J] = i \delta^{IJ}~.
\end{equation}
These can be used to show, for example, that
\begin{eqnarray}\label{chidchicom}
[\chi^a(\bx;U), \pd_M \chi^b(\by;U) ] &\approx& 0\cr
 [\pi^a(\bx;U),\pd_M \pi^b(\by;U)]&\approx& 0 ~. 
\end{eqnarray}
More generally, the commutator of any $U$-derivative of $\chi$ with
another $U$-derivative of $\chi$ is zero, and similarly for $\pi$.

\section{Some details on the canonical
  transformation}\label{app:commutators}

First let us consider $\{ \Phi(\bx), \Pi(\by) \}_{\rm D}'$ in order to
derive the classical form of $\Pi_{0}^M$ as given in
\eqref{classicalPi0}.  Substituting in \eqref{cov}, \eqref{cov2} for
$\Phi,\Pi$ and using \eqref{DBs2}, we can write the result as
\begin{align}\label{PhiPi1}
  \{ \Phi^a(\bx), \Pi^b(\by) \}_{\rm D}' \approx&~ \delta^{ab}
  \delta(\bx-\by) + \pd_M \phi^a(\bx) \left( \frac{\pd \Pi_{0}^N}{\pd
      p_M} - G^{MN} \right) \pd_N \phi^b(\by) \cr & - \pd_Q
  \chi^a(\bx) \frac{\pd \Pi_{0}^N}{\pd p_M} \pd_N \phi^b(\by)  \cr &
  + \int \ed z \left( \delta^{ac} \delta(\bx-\bz) - \pd_M \phi^a(\bx)
    G^{MQ} \pd_Q \phi^c(\bz) \right) \frac{\delta \Pi_{0}^N}{\delta
    \pi^c(\bz)} \pd_N \phi^b(\by) ~. \qquad \quad
\end{align}
The first term is what we want; thus we must choose the functional
$\Pi_{0}^M$ so that the remaining terms vanish.  Consider the $\bx$
dependence of these remaining terms.  The term in the first line is
tangential to $T_U \MM \subset L^2[\mathbb{R}_{(\bx)}^{D-1}]$ since it
is proportional to the zero-mode $\pd_M \phi(\bx)$, while the term in
the last line is in the orthogonal complement $(T_U \MM)^\perp$ since
it involves the projection operator $\delta(\bx-\bz) - \pd_M \phi(\bx)
G^{MQ} \pd_Q \phi(\bz)$.  The term involving $\pd_Q \chi(\bx)$ can be
decomposed into a piece along $T_U \MM$ and a piece orthogonal to it.
Substituting this into \eqref{PhiPi1}, we find that $\{
\Phi(\bx), \Pi(\by) \}_{\rm D}' = \delta(\bx - \by)$ if and only if
both of the following hold:
\begin{align}\label{vanish1}
  0 \approx&~ \frac{\pd \Pi_{0}^N}{\pd p_M} - G^{MN} - \frac{\pd
    \Pi_{0}^N}{\pd p_M} G^{MP} \int \pd_P \phi \cdot \pd_Q \chi ~, \cr
  0 \approx&~ \pd_Q \chi^c(\bz) \frac{\pd \Pi_{0}^N}{\pd p_M} + \frac{\delta
    \Pi_{0}^N}{\delta \pi^c(\bz)}~.
\end{align}
Note that we can write
\begin{equation}
  \int \pd_P \phi \cdot \pd_Q \chi = \pd_Q \psi_{P}^{(1)} - \int \chi
  \cdot \pd_P \pd_Q \phi \approx - \int \chi \cdot \pd_P \pd_Q \phi
  = -  \Xi_{PQ}~, 
\end{equation}
where $\Xi_{PQ}$ was defined in \eqref{Xidef}.  The first of
\eqref{vanish1} implies
\begin{equation}
\frac{\pd \Pi_{0}^N}{\pd p_M} \approx \left[ (G - \Xi)^{-1} \right]^{MN} =: C^{MN}
\end{equation}
whence the second equation implies that
\begin{equation}\label{Pi0result}
\Pi_{0}^N \approx \left( p_M - \int \pi \cdot \pd_M \chi \right) C^{MN}~.
\end{equation}
Here we have omitted the possible addition of a term depending only on
the coordinates $(U,\chi)$.  Consideration of $\{ \Pi(\bx), \Pi(\by)
\}_{\rm D}'$ shows that it is consistent to set this term to zero.

We observe that if we set $\chi, \pi =0$, then the momentum
transformation \eqref{cov2} with \eqref{Pi0result} reduces to
$\Pi(\bx) = p^M \pd_M \phi(\bx;U)$.  This is exactly what one would
expect for the classical momentum density of the moving soliton.

At the quantum level we take the change of momentum variables to be
\eqref{quantumPi0}.  The basic commutators are the right-hand sides of
\eqref{DBs2}, multiplied by a factor of $i$.  Using these we have
\begin{align}\label{chiacom}
  & [ f(U), a^M]' = [f(U), \bar{a}^M]' = i C^{MN} \pd_N f(U)~, \cr &
  [\chi(\bx;U), a^M]' = [\chi(\bx;U), \bar{a}^M]' \approx i \left(
    G^{MN} - C^{MN} \right) \pd_N \phi(\bx;U)~,
\end{align}
where $f$ is any function of $U$.  Then one easily obtains the desired relation,
\begin{equation}
[\Phi^a(\bx), \Pi^b(\by)]' \approx i \delta^{ab} \delta(\bx - \by) ~.
\end{equation}

For $[\Pi(\bx),\Pi(\by)]'$ we first note that
\begin{equation}
[\pi(\bx), C^{MN}] = -i C^{MP} \left( \nabla_P \pd_Q \phi(\bx) \right) C^{QN}~,
\end{equation}
from which it follows that
\begin{align}\label{piacom}
  & [\pi(\bx), a^M]' \approx - i \Theta_{PN} C^{NM} G^{PQ} \pd_Q
  \phi(\bx) - i a^N C^{MP} \nabla_N \pd_P \phi(\bx)~, \cr & [\pi(\bx),
  \bar{a}^M]' \approx -i \left( \nabla_P \pd_N \phi(\bx) \right)
  C^{MP} \bar{a}^N - i \left( \pd_Q \phi(\bx) \right) G^{QP} C^{MN}
  \Theta_{NP}
\end{align}
and where we have defined
\begin{equation}
\Theta_{MN} := \int \pi\cdot \pd_M \pd_N \phi~.
\end{equation}
Note that  $\Theta_{MN}  = \Theta_{(MN)}$.   Using  this,  one can  express
$[\Pi,\Pi]$ in the form
\begin{align}\label{PiPicom1}
[\Pi^a(\bx), \Pi^b(\by)]' \approx&~ - i a^P C^{Q[M}
\Gamma^{N]}_{\phantom{N}PQ} \pd_M \phi^a(\bx) \pd_N \phi^b(\by) + i
\pd_M \phi^a(\bx) \pd_N \phi^b(\by) \Gamma^{[M}_{\phantom{M}PQ}
C^{N]P} \bar{a}^Q  \cr 
& + i \left( C \Theta G^{-1} - G^{-1} \Theta C \right)^{[MN]} \pd_M
\phi^a(\bx) \pd_N \phi^b(\by)  \cr & + \frac{1}{4} [a^M, a^N] \pd_M
\phi^a(\bx) \pd_N \phi^b(\by) + \frac{1}{4} \pd_M \phi^a(\bx) \pd_N
\phi^b(\by) [\bar{a}^M, \bar{a}^N]  \cr & + \frac{1}{4} \pd_N
\phi^b(\by) [a^M, \bar{a}^N] \pd_M \phi^a(\bx) - \frac{1}{4} \pd_N
\phi^a(\bx) [a^M, \bar{a}^N] \pd_M \phi^b(\by)~. \raisetag{16pt}
\end{align}

What one needs is then the commutators of the $a$'s and $\bar{a}$'s.
Equations \eqref{chiacom} and \eqref{piacom} together with
\begin{align} [p_Q, C^{MN}] =&~ -C^{MR} [p_Q, G_{RS} - \Xi_{RS}]
  C^{SN} = i C^{MR} \left( \pd_Q G_{RS} - \pd_Q \Xi_{RS} \right)
  C^{SN}
\end{align}
can be used to show
\begin{align}
  & \left[C^{PM}, p_Q - \smallint \pi \cdot \pd_Q \chi \right] =
  \left[C^{PM}, p_Q - \smallint \pd_Q \chi \cdot \pi \right] \approx\cr &
  \qquad \qquad \qquad \qquad \approx -i C^{PR} C^{MS} S_{QRS} + i
  C^{PR} C^{MS} \Gamma^{T}_{\phantom{T}RS} (C^{-1})_{TQ}~, \qquad
  \quad
\end{align}
where we have defined
\begin{equation}
  S_{QRS} := \pd_Q G_{RS} + \Gamma_{QRS} - \smallint \chi \cdot \pd_Q
  \pd_R \pd_S \phi~, 
\end{equation}
which is totally symmetric, $S_{QRS} = S_{(QRS)}$.  Making note of the
comment below \eqref{chidchicom} and using \eqref{dchipi}, one also
finds that
\begin{align}
  \left[ (p_P - \smallint \pi \cdot \pd_P \chi), (p_Q - \smallint \pi
    \cdot \pd_Q \chi) \right] \approx &~ -2 i \Theta_{[P|R} G^{RS}
  (C^{-1})_{S|Q]} ~, \cr \left[ (p_P - \smallint \pd_P \chi \cdot
    \pi), (p_Q - \smallint \pd_Q \chi \cdot \pi) \right] \approx &~ 2
  i (C^{-1})_{[P|R} G^{RS} \Theta_{S|Q]} ~, \cr \left[ (p_P -
    \smallint \pi \cdot \pd_P \chi), (p_Q - \smallint \pd_Q \chi \cdot
    \pi) \right] \approx &~ i \Theta_{QR} G^{RS} (C^{-1})_{SP} - i
  (C^{-1})_{QR} G^{RS} \Theta_{SP}  \cr & + i \int [\pd_Q \pd_P
  \chi^a, \pi^a]  \cr & - \int \ed\bz \ed\bw [\pd_Q
  \chi^b(\bw),\pi^a(\bz)] [\pd_P \chi^a(\bz), \pi^b(\bw)] ~. \qquad
  \quad
\end{align}
These imply
\begin{align}\label{aacom}
  & [a^M,a^N] \approx 2i \left( a^P C^{Q[ M}
    \Gamma^{N]}_{\phantom{N}PQ} + (G^{-1} \Theta C)^{[MN]} \right)
  \cr & [\bar{a}^M, \bar{a}^N ] \approx -2i \left(
    \Gamma^{[M}_{\phantom{M}PQ} C^{N]P} \bar{a}^Q + (C \Theta
    G^{-1})^{[MN]} \right)
\end{align}
and
\begin{align}\label{aabarcom}
[a^M, \bar{a}^N]  \approx &~ i a^{P} C^{QM} \Gamma^{N}_{\phantom{N}PQ}
- i \Gamma^{M}_{\phantom{M}PQ} C^{NP} \bar{a}^Q + i \left( C \Theta
  G^{-1} - G^{-1} \Theta C \right)^{NM}  \cr 
&~ - i a^P C^{MQ} C^{NR} S_{PQR} + i S_{PQR} C^{MP} C^{NQ} \bar{a}^R 
\cr &~ - C^{NQ} \bigg\{ \Gamma^{R}_{\phantom{R}SP}
\Gamma^{S}_{\phantom{S}RQ} + C^{RS} C^{TV} S_{RTP} S_{SVQ} - 2
\Gamma^{R}_{\phantom{R}(P|S} C^{ST} S_{TR|Q)}  \cr & \qquad \quad -
\int [\pd_Q \pd_P \chi^a, \pi^a] + \int \ed\bz \ed\bw [\pd_Q
\chi^b(\bw),\pi^a(\bz)] [\pd_P \chi^a(\bz), \pi^b(\bw)] \bigg\} C^{MP}
~. \quad \cr
\end{align}

When substituting \eqref{aabarcom} into \eqref{PiPicom1}, the last two
lines of \eqref{aabarcom} do not contribute because they commute with
$\pd_M \phi$ and are symmetric in $M,N$.  Furthermore, after commuting
all $a$'s to the far left and all $\bar{a}$'s to the far right, and
using the symmetry properties of $C^{MN},S_{MNP}$, there are
additional cancellations and one is left with
\begin{align}\label{aabarcom2}
  & \pd_N \phi^b(\by) [a^M, \bar{a}^N] \pd_M \phi^a(\bx) - \pd_N
  \phi^a(\bx) [a^M, \bar{a}^N] \pd_M \phi^b(\by) \approx \cr & \qquad \qquad
  \approx 2i a^{P} C^{Q[M} \Gamma^{N]}_{\phantom{N}PQ} \pd_M
  \phi^a(\bx) \pd_N \phi^b(\by) - 2i \pd_M \phi^a(\bx) \pd_N
  \phi^b(\by) \Gamma^{[M}_{\phantom{M}PQ} C^{N]P} \bar{a}^Q \cr &
  \qquad \qquad \quad +2i \left( G^{-1} \Theta C - C \Theta G^{-1}
  \right)^{[MN]} \pd_M \phi^a(\bx) \pd_N \phi^b(\by) ~.
\end{align}
Using \eqref{aacom} and \eqref{aabarcom2} in the calculation of
\eqref{PiPicom1} leads to complete cancellation on the constraint
surface:
\begin{equation}
[ \Pi^a(\bx), \Pi^b(\by) ]' \approx 0~.
\end{equation}
%


\bibliographystyle{utphys}
\bibliography{reduction}

\end{document}